# Phase mapping of aging process in InN nanostructures: oxygen incorporation and the role of the zincblende phase

D. González,[1] J. G. Lozano[1], M. Herrera[1], F. M. Morales[1], S. Ruffenach[2], O. Briot[2] and R. García[1]

[1] Departamento de Ciencia de los Materiales e I. M. y Q. I., Universidad de Cádiz, Puerto Real, Cádiz, Spain.
[2] Groupe d'Etudes des Semiconducteurs, , Université Montpellier II UMR 5650 CNRS Place Eugène Bataillon, 34095 Montpellier, (France).
E-mail: david.gonzalez@uca.es

**Abstract.** Uncapped InN nanostructures undergo a deleterious natural aging process at ambient conditions by oxygen incorporation. The phases involved in this process and their localization is mapped by Transmission Electron Microscopy (TEM) related techniques. The parent wurtzite InN (InN-w) phase disappears from the surface and gradually forms a highly textured cubic layer that completely wraps up a InN-w nucleus which still remains from original single-crystalline quantum dots. The good reticular relationships between the different crystals generate low misfit strains and explain the apparent easiness for phase transformations at room temperature and pressure conditions, but also disable the classical methods to identify phases and grains from TEM images. The application of the geometrical phase algorithm in order to form numerical moiré mappings, and RGB multilayered image reconstructions allows to discern among the different phases and grains formed inside these nanostructures. Samples aged for shorter times reveal the presence of metastable InN:O zincblende (zb) volumes, which acts as the intermediate phase between the initial InN-w and the most stable cubic $In_2O_3$ end phase. These cubic phases are highly twinned with a proportion of 50:50 between both orientations. We suggest that the existence of the intermediate InN:O-zb phase should be seriously considered to understand the reason of the widely scattered reported fundamental properties of thought to be InN-w, as its bandgap or superconductivity.

## 1. Introduction

Indium nitride (InN) has been intensively studied in recent years because of unusual unexpected characteristics such as small effective masses, large electron drift velocities, high electron accumulations at their surface layers and their bandgaps [1]. This has opened many device application possibilities up such as emitter/detectors in the infrared range, solar cells, THz surface emitters, etc. Recently, low dimensional nanostructures of GaN and InN, such as such as nanoislands, nanowires, nanorods, nanobelts and nanotubes, have received a growing interest due to their novel size- and dimension-dependent physical properties.[2,3] In particular, its unusual surface properties and inherently heavily doped nature highlight among them and they should, in fact, be exploited for uses in microelectronic devices.

Yet there have been huge efforts to prepare InN nanostructures, the reported results are not altogether complete and satisfactory.[4] The reasons must be found in the difficulties both in the growth of high quality structures and in the application of standard measurement techniques developed for other III–V materials.[5] As a consequence, InN has still generated several controversies concerning its fundamental parameters. As an example, the disparity about the real bandgap of the InN seems to find no end. In earlier years, the bandgap of InN seemed to be settled at a value of 1.89 eV.[6] However, in the last decade the reported bandgap values have progressively dropped to lower ones (1.1 eV,[7] 0.90 eV,[8] 0.70 eV,[9] down to the range 0.65–0.60 eV).[10] Nevertheless, there is a lately trend to consider higher values of the InN bandgap in the range of 1.1–1.7 eV.[11-13] It is clear that a new view to shed light on the understanding of this broad dispersion of data is needed and nowadays nobody neglects the role of oxygen to provide explanations about this topic. In fact, many of the variations in the measured values of the bandgap are attributed to oxygen contamination, either by the formation of semiconducting pseudo-binary alloy phases or by impurity banding within the band-gap.[14,15]

On the other hand, a similar controversy related to the occurrence of superconductivity in some epitaxially grown InN layers is raising.[16,17] Several authors observed a quick drop in the electrical resistivity below 4.2 K in InN layers and proposed that neither the surface electron accumulation layer nor the metal-In precipitation contribute to this superconductivity state.[18,19] This strong change in the resistivity has been suggested as the sign of a phase transition to a possibly superconducting state. However, Kadir et al.[20] have recently established that the superconductivity is not intrinsic to the InN layers since no superconductivity is found in samples where traces of indium oxide are not observed and it arises more as the oxygen content increases. In spite of the fact that oxygen plays an important role in defining the overall properties of InN structures, a detailed study of the structural properties of the oxidized InN nanostructures has not yet been made.

In this work, we show the aging process of InN nanostructures due to atmospheric exposure, which takes place through the chemiabsorption of oxygen atoms into the InN. This route implies the gradual transformation of the In sublattice from the hexagonal arrangement in wurtzite InN to the cubic stacking in stable $In_2O_3$, where metastable zincblende InN domains rich in oxygen atoms are proposed to act as intermediate phases. On this purpose, a methodology to obtain phase mappings of this transformation in InN nanostructures is anticipated by applying geometric phase (GP) algorithms [21,22] and RGB multilayered reconstructions to high-resolution TEM (HRTEM) images. The aim is to provide useful tools to characterize the quality of the monolayers very close to surface in InN nanostructures.

## 2. Experimental setup

InN nanostructures samples were grown on $GaN/Al_2O_3$ by Metalorganic Vapor Phase Epitaxy. The GaN buffer layer was grown on a (0001) sapphire wafer using the usual process at a temperature close to 1000°C in all cases. The temperature was then lowered to 550°C and the InN nanostructures were deposited using Trimethylindium and ammonia as In and N precursors, respectively. Sample A is an uncapped InN nanostructure structure grown 36 months before this study and maintained under room temperature and pressure conditions. These InN dots were grown under a V(In)/III(N) molar ratio of 5000 at 550°C. Sample B was grown by using a V/III ratio of 15000 with an average height of 12±2 nm. The growth time and the V/III molar ratio were used as control parameters to tune the size and density of the InN quantum dots.[23] The dots of sample B were grown to maximize the areal density of dots and enabled to increase the InN dot density reducing the size by a factor of 5 compared to the elder sample A. A complete characterization of the morphology and relaxation state can be found elsewhere.[24,25]

The HRTEM images and electron energy loss spectroscopy (EELS) in scanning mode were carried out in cross section geometry samples using JEOL 2010F and JEOL JEM 2500SE microscopes, both working at 200kV. The image processing and the GP algorithms were performed using routines written for Matlab 7.1.[26] All the FFT images were built using the logarithm of the absolute value of the complex data array obtained from the FFT of the 16-bit HRTEM images. The FFT images are a representation of theses matrixes in a 256 greyscale. Considering the centre symmetry of the FFT images when the input data are purely real, we operate with half of the FFT reconstruction. In order to analyze the error attributed to the noise in the assignment of the peaks in the FFT, a signal to noise ratio (*SNR*) analysis was performed using the industry standard ISO 12232: Thus, the *SNR* in decibels (dB) of power is defined as:

$$SNR(dB) = 20\log\left(\frac{Signal}{RMS\ noise}\right) \quad (1)$$

where the signal is the average intensity in the peak minus the background values and the *RMS* or root mean square noise is defined from the background region, namely, the square root of the absolute value of the sum of variances. For instance, a ratio of 1:1 yields 0 decibels. Following this procedure, we consider as strong spots those with a *SNR* value higher than 20 dB and weak spots are those ones ranging between 10 to 20 dB.

Selective mask areas and RGB multilayered reconstructions of the HRTEM images of the nanoislands were performed by using the Gatan *Digital Micrograph®* software. Electron diffraction patterns based on the dynamical theory were calculated by the *EMS* software (P. H. Jouneau and Pierre Stadelmann, EPFL, Lausanne).

3. **TEM results.**

**Figure 1.a** corresponds to a HRTEM image taken along the $[11\bar{2}0]$ zone axis of an uncapped InN nanoisland of sample A that was studied just after being grown. The (0001) planes follow a stacking sequence of a hexagonal structure, namely *ABAB*, and this fact is reflected in the Fast Fourier Transform (FFT) of the image. Certainly, FFT images provide information about the periodicity of the atomic structure and they can be correlated to the electron diffraction patterns registered at the back focal plane of the objective lens in a TEM microscope.[27] In the following, we assume the bright spots in the FFT image as diffraction spots (reflections) from crystallographic planes of the crystalline phases in the structures and taking into account the symmetry of the FFT images when the input data are purely real, we do not draw the symbols that have their corresponding symmetric mirror image regarding the centre of the FFT reconstruction. In this manner, the **Fig. 1.b** shows the FFT pattern where spots correspond to a diffraction pattern of pure wurtzite structures of GaN-w and InN-w in the $[11\bar{2}0]$ zone axis. These results are the same to all the fresher samples, that is to say, only the wurtzite phases are present and they do not contain any appreciable volume of cubic phases in the GaN or InN regions.

After these studies, the InN nanostructures were stored under room conditions into boxes and were re-examined again after different times of air exposure. **Figure 2** shows HRTEM images of InN nanostructures taken along the $[11\bar{2}0]$ zone axis of sample B (a) and sample A (b), respectively (studies of cross sectional TEM preparations were performed after 36 months for sample A and after 12 months for sample B). An alteration of the stacking sequence in specific regions of the InN nanostructures is observed and this implies a change in the FFT pattern. **Figure 3** includes the FFT images of the HRTEM micrographs presented in **Figure 2**. In this way, in **Figure 3(a)** the spots marked by yellow triangles correspond to a diffraction pattern of an InN-w crystal registered along the $[11\bar{2}0]$ zone axis, being the (0002) spot associated to those planes parallel to the growth direction. The extra spots in the FFT surrounded by green and red squares correspond to the diffractograms of two InN-zb crystals rotated 180° around the growth direction, respectively. Both crystals have {111} spots aligned along the growth direction which coincide spatially with the {0002} spots related to the InN-w. This kind of pattern is typical for the less aged samples (from 3 months onwards) and implies an intermediate step in the gradual transformation of the Indium sublattice from the hexagonal stacking proper of InN to that of cubic oxidized structures.[28] Although metastable InN doped with oxygen (InN:O) has been proposed as an intermediate phase between the InN-w and the more thermodynamically stable bcc-$In_2O_3$ [29-32 phase, it has not been structurally characterized. Nevertheless, this pattern changes with the time of aging as shown in the FFT image of the elder sample (sample A) displayed in **Figure 3(b)**. This pattern acquires a higher level of complexity because new spots arise. Similar to the FFT pattern of the InN:O-zb phase, the new spots corresponds to that of two twined $In_2O_3$ crystals for which the {222} spots match with the (0002) spot of InN-w.[33] Even more, many spots share their positions in the reciprocal space with the spots of the InN:O-zb phase in this case. For a better visualization of this coincidence, **Figures 4(a)** and **4(b)** show the calculated dynamical diffraction patterns in the [110] zone axes of InN-zb and $In_2O_3$ phases where black spots have their sizes proportional to the expected intensity of diffraction, and red arrows are aligned along the growth direction. If overlapped, many of the spots in the calculated patterns will coincide and it is then easy to understand that in the experimental FFT image displayed in the **Figure 4(c)** all the spots in the InN-zb pattern coincides in the reciprocal space with the stronger reflections of the $In_2O_3$ pattern. Note that the direction of the red arrow in this FFT pattern is equivalent to those shown in the calculated diffraction patterns. No additional spots are visible, such as those that might occur from oxidized GaN phases. The stability from air exposure of the GaN-w surfaces is much higher compared to InN-w surfaces, at least at the same conditions.[28]

From this analysis, it is deduced that aged InN nanostructures contain the primitive InN-w together with two new phases, InN:O-zb and $In_2O_3$, both in two different orientations. In this way, the FFT pattern offers information about five crystal orientations inside the nanoisland but, in spite of this, the number of significant spots in the pattern is not very large. This large quantity of spots coincidences in the reciprocal space implies to a good concurrence in orientations and interplanar spacings among the different crystals due to a topotactic reaction. Thus, the position of the (0002) spots of the InN-w in the reciprocal space coincides with the {111} spots of both twinned InN:O-zb crystals and with both {222} spots of the $In_2O_3$ crystals (i. e. the interplanar distances for planes parallel to the substrate surface are the same in the five kind of crystals). There are more coincidences among the interplanar distances and this fact explains the sharp column contrast observed in the HRTEM images and the conservation of the morphology and size of the nanostructures during the aging process.

Nonetheless, the lack of mismatch-related defects inside the nanoisland, as dislocations or high-angle boundaries is an impediment to obtain direct information about the location of the different phases inside the nanoislands. The commented good reticular relationships between the different crystals generate low misfit strains and explain the apparent easiness for phase transformations at room temperature and pressure conditions, but also disable the classical methods to identify phases and grains from TEM images. Moreover, chemical analytical methods as electron energy loss spectroscopy (EELS) in scanning TEM mode could be applied to reveal the composition, but it hardly revealed data about the crystallography or the microphases presented. Anyway, the presence of oxygen has been checked in a series of core loss EEL spectra along the [0001] axis of an aged InN nanoisland from the interface to the surface. **Figure 5** displays the K edge for nitrogen (401 eV), $M_{4,5}$ edges for indium (451 eV) and K edge for oxygen (532 eV) for three different spatial positions of the heterostructure shown in Fig. 2b. We can observe the decay of the nitrogen edge as well as the raise of the oxygen edge when approaching the surface and this fact implies the progressive substitution of N by O atoms in the direction to the surface. Although the oxygen incorporation is revealed, the phases and the texture during the agedness of the InN nanostructures must be solved. In the following, a developed methodology for HRTEM image processing is implemented in order to reconstruct with precision the crystal positions and to visualize the phase mapping of the oxidation process in InN nanostructures.

## 4. Phase mapping

The geometric phase or GP method is assiduously applied to obtain deformation maps of the lattice displacements and strain fields from HRTEM images. In a few words, the method consists of constructing a phase map for a given Bragg condition with respect to an ideal reference lattice. For this, it is necessary to center small apertures (Bragg masks) on two strong reflections of the FFT pattern of an HRTEM image and to perform a subsequent inverse Fourier transform. A full description of the methodology can be found in Refs 21 and 22. However, in this work the geometrical phase algorithm is not used with the usual purpose, but we have used it to build a methodology that allows us to obtain maps of the location of the different phases inside the nanostructure. For this, we built numerical moiré images, $M(\mathbf{r})$, superimposing the real lattice with a reciprocal lattice vector smaller than the average lattice where $M$ is a magnification constant defined as [34]

$$M(\mathbf{r}) = \frac{2\pi \, \mathbf{g}_r \cdot \mathbf{r}}{M} - 2\pi \, \mathbf{g}_r \cdot \mathbf{u}(\mathbf{r}) \qquad (2)$$

where $\mathbf{g_r}$ is the reference lattice in the reciprocal space and $\mathbf{u}(\mathbf{r})$ is the displacement of the atomic column position from its perfect position. Technically, this is carried out by changing the origin for the reference lattice in the FFT image to the point $|\vec{\mathbf{g}}|/M$.[35] The moiré pattern acts as a lens which magnifies not only the lattice mismatch but also the distortions and rotations by a factor $M$.[36]

Following this procedure, the numerical moiré images were obtained using as reference $\mathbf{g_r}$ the position of the spot $(\bar{1}1\bar{2})$ in the $In_2O_3$ pattern and using a Bragg mask that includes the stronger $\{hhh\}$ spots of cubic phases and the $\{01\bar{1}0\}$ of the hexagonal phases (see blue circle in **Fig. 4(a)**). In order to analyze the error attributed to the noise in the assignment of the peaks in the FFT, and therefore the reliability of the analyzed reflections, a signal to noise ratio ($SNR$) analysis was performed. For comparison, the $\{hhh\}$ spots have a $SNR$ of 21.6 dB and the $\{\bar{1}12\}$ spot a $SNR$ of 17.8 dB. With the mentioned Bragg mask radius, we gather the information from all the crystals that appear in the structure. Following this procedure, **Figure 6** shows the numerical moiré images obtained from the HRTEM micrographs shown in **Figure 2(a)** and **2(b)**, respectively. They reveal four different sets of lattice fringes that correspond to the four FFT spots included inside the mask. Thus, there are two sets of vertical fringes where the **g** vectors are aligned with the reference $\mathbf{g_r}$ yielding a translational moiré pattern. These regions are consistent with the presence of wurtzite phases where the moiré spacing is proportional to $1/(\mathbf{g}\text{-}\mathbf{g_r})$. So, the nearer the **g** vectors stand, the wider the moiré spacings are. As a consequence, the narrower and the wider vertical fringe patterns can be associated with the $\{01\bar{1}0\}$ spots of the wurtzite phase from the GaN-w substrate and the InN-w, respectively. On the other hand, the regions near the surface present two sets of inclined fringes that correspond to rotational moiré patterns coming from the interference with the {$hhh$} spots of cubic phases included in the mask. The regions in the **Figure 6** with moiré fringes inclined to the right " \\\ " come from the interference with the spots in the upper region in the mask, and the ones inclined to the left "///", with the spots in the lower region. In the **Figure 6(a)**, the nanoisland is in the beginning of the oxidation process, where the spots characteristics of the $In_2O_3$ do not appear but the intermediate InN:O-zb doped with oxygen related spots do. Only, a clean grain of cubic phase near the nanoisland tip and an incipient blurred thin layer at the surface containing both orientations can be observed. The elder sample, **Figure 6(b),** clearly shows that the InN-w phase has disappeared in the regions close to the surface shrinking to the centre of the nanoisland and thus reducing its average thickness from 12 to 6 nm. This loss of hexagonal phase in the pyramidal facets is even higher, giving an average thickness of ~6.5 nm. The cubic phases form a layer that envelops completely the core of InN-w. In addition, the FFT image points the presence of a twin pattern for the cubic phases out, and the images of numerical moiré patterns present the resolved spatial location of these crystals. The nanostructure of the oxidized InN layer consists of an alternating sequence of twinned domains of cubic phases where the proportion between the two orientations is around 50:50 in the majority of the eldest nanostructures studied. The natural aging process transforms the wurtzite phase of the parent InN into a polycrystalline cubic structure originated at the surface.

However, so far we could not separate in this case the InN:O-zb phase from the $In_2O_3$ phase. The numerical moiré images draw the contribution of both phases since they share similar spatial positions in the masked FFT region of the reciprocal space. Certainly, the major contribution in the intensity of these spots likely comes from the $In_2O_3$ phase but we cannot choose a reflection of the InN:O-zb phase without the input of the $In_2O_3$ phase (see **Fig. 4**). Fortunately, the complexity of the unit cell of the $In_2O_3$ causes noticeable spots further from the main reflections of the InN:O-zb and there are extra spots in the FFT pattern that can be unequivocally assigned to each orientation of the indium oxide, namely, the {114}, {112} and {333} families (see **Fig 4.b**). However, the intensity of these spots is weak (*SNR* ranges from 10 to 20) and to apply the GP algorithm to construct numerical moiré mappings with statistically significant results was not possible. Moreover, although the relative intensity of non-shared spots from both $In_2O_3$ patterns is feeble, we can collect the information on a high number of distinctive spots for each orientation. Taking into account this fact, we have constructed two selective Bragg mask patterns collecting independently the separated contribution of each $In_2O_3$ crystal that do not share positions with the InN:O-zb related spots. **Figure 7** shows the RGB multilayered reconstruction from three inverse FFT (iFFT) images of these selected masks representing the red, green and blue components. The red areas correspond to the iFFT image of the Bragg mask pattern for red-circled spots in the **Figure 3.b** that do not coincide with any other reflection. The same procedure was used for the green component with the non shared green circle spots, whereas the blue component is the blank. The figure clearly shows an $In_2O_3$ surrounding layer that envelops the InN-w nucleus, but in a discontinuous way. The proportion for both twinned components is around 50:50 following an alternative sequence separated by regions of cubic oxidized phases of InN:O-zb.

Additionally, the structural relation of the hexagonal InN-w with cubic $In_2O_3$ and InN:O-zb phases in the real space can be deduced from **Fig. 6(b).** The parent InN-w phase is reduced to a nucleus inside the nanostructures with a truncated pyramid shape. Thus, the top facet of InN-w (0002) is parallel to the cubic $In_2O_3$ {222} and InN-zb {111} facets forming habit planes inside the nanostructure. The pyramidal facets on the left- and right-hand side of the hexagonal InN-w nucleus are mainly of $\{10\bar{1}n\}$ type (being *n*=1,2) and they have an orientation relationship with the cubic $In_2O_3$ $(22\bar{2})$ and $In_2O_3$ (004) facets. Indeed, the hexagonal InN-w $(01\bar{1}1)$ would be placed in a ~7°-twin state to the cubic $In_2O_3$ $(00\bar{4})$/InN-zb $(00\bar{2})$ facet and in the ~10°-twin state to the cubic $In_2O_3$ $(2\overline{22})$/InN-zb $(1\overline{1}\overline{1})$.[37] In general, the diffractograms of all phases are well aligned regarding the growth direction although residual rotations between the cubic and the hexagonal phases are sometimes hinted..

## 5. Discussion

The numerical moiré fringes mapping of **Figure 6(b)** shows a continuous shield of cubic phases that wraps up a residual InN-w nucleus but the nanostructure dimensions are well retained during the oxidation of InN to $In_2O_3$. This observation can be explained by a shrinking core model [38] as the reaction route, in which nitrogen will be substituted by oxygen and the structure changes from the wurtzite structure of InN-w to the cubic structure of bcc- $In_2O_3$. The oxidation of the initial InN nanoisland starts at the surface and during growth of this oxide layer, the inner InN nucleus seems to shrink. This model assumes that the transport of oxygen and nitrogen is not determining the rate, instead, the proper reaction at the interface between the InN and the Indium oxide phase is the rate determining step. Under this assumption, the reaction rate is proportional to the area and this fact supposes a drawback for the fabrication of InN based nanostructures since the surface-to-volume ratio is high.

What is more, the huge difference between their formation enthalpy (-131.9 kJ/mol for InN-w and -925.3 kJ/mol for $In_2O_3$) [[39,40] and the low reaction temperature (room conditions) imply a high chemical free energy (negative) for the nucleus formation and a low diffusion coefficient, witch means a quicker nucleation rate and a slower growth rate and, as a result, many small nuclei are formed. Thus, **Figure 6.b** and **Figure 7** show that Indium oxide phases replace the InN-w phase by forming several twined domains smaller than 5 nm in diameter. In addition, the shear processes involved in the transformation also promote a highly twinned structure. In the very early stages of the oxidation reaction these nuclei form on the surface of the InN structure, and once they have grown together, a dense oxide layer develops. Additionally, the nucleation rate in the oxidation reaction, which will in turn depend upon the availability of nucleation sites, may proceed rapidly in defective samples or in high specific surface area structures such as the nanostructures.

Finally, we point out the existence of InN-zb regions or, at least, InN:O-zb (InN-zb doped with oxygen). For the III-N group of compounds, the total energy of the wurtzite phase is smaller than that of the cubic one, i.e., the wurtzite form is more stable than the zincblende form.[41] However, theoretical predictions presume that this difference for InN is very small (17 eV/unit cell) [42] and metastable zincblende InN can appear. These kinds of metastable structures are particularly prevalent in covalently bonded materials, reflecting the slow atomic self-diffusion in materials where bond formations and breakings are involved.[43] In addition, the ending product, $In_2O_3$, has a bixbyite structure with the space group $Ia\overline{3}$ and we can describe as a cubic stacking sequence *ABCA'B'C*. Ideally, six oxygen atoms surround each indium atom although the existence 16c vacant anion sites characteristics of the bixbyite structure of $In_2O_3$ favors the formation of intermediate structures.[44]

As a consequence, when oxygen atoms gradually replace nitrogen atoms, it is easy to understand a continuous geometrical transformation from InN-w towards $In_2O_3$. During the natural aging of InN, the In sublattice is in transit from a *ABAB* stacking sequence (InN-w) towards a *ABCA'B'C'* stacking ($In_2O_3$).[28] The formation of an intermediate InN:O-zb with a *ABCABC* stacking easily explains the intermediate steps necessary for this transformation. Thus, a mechanism for the natural aging can be proposed: In a fist step, a progressive chemical adsorption of the atmospheric oxygen occurs at the surface. Later, the Indium sublattice changes from the *ABAB* stacking to the *ABCABC* one giving place to InN:O alloys with a zincblende structure $(F\overline{4}3m)$. This process implies a homogeneous shear of basal planes of the InN-w phase that would also explain the highly twinned microstructure of the oxide phases. In a further state, when the oxygen proportion arises decreasing the nitrogen content, the Indium sublattice change/lose its symmetry changing to a *ABCA'B'C'* stacking corresponding to a bixbyite structure. Further work is in progress to analyse the aging process and its temporal evolution in flat layers. In this way, we suggest that in order to understand the fundamental properties of InN, as the bandgap or the superconductivity, the existence of the described intermediate phase InN:O-zb should be seriously considered.

## 6. Conclusions

In summary, InN nanostructures suffer a phase transformation by their exposure to air at room temperature and pressure conditions. The FFT of cross sectional HRTEM images of samples at different aging times showed the presence of cubic phases together with the parent InN-w. Although $In_2O_3$ is the main component of the cubic regions in the elder samples, the presence of InN:O with a zincblende structure is also detected. We have developed HRTEM methodologies to obtain maps of the spatial locations of the different phases and grains. Firstly, numerical moiré fringes mappings obtained from the application of the geometrical phase algorithm to different micrographs show that cubic phases substitute the InN-w parent phase forming a surrounding layer that envelopes completely a core of the original InN-w for the elder samples. Secondly, RGB multilayered reconstructions using selective masks evinced the formation of a metastable InN:O-zb phase acting as intermediate phase between the parent phase of InN-w and the ultimate phase of $In_2O_3$. These cubic phases are highly twinned with a proportion of 50:50 between both orientations.
.

**Acknowledgments**
The authors are grateful to Dr. Pedro Galindo for the interesting and fruitful discussions. Financial support from CICYT project MAT2007-60643 (Spain) and SANDIE European Network of Excellence (NMP4-CT-2004-500101—Sixth Framework Program) is gratefully acknowledged.

**Figure captions**

**Figure 1.** (a) Cross sectional HRTEM image taken along the $[11\bar{2}0]$ pole of the uncapped sample A shortly after being grown. (b) FFT of Fig.1(a) where no hint of cubic phase is present.

**Figure 2**. X-HRTEM micrographs of uncapped InN nanostructures taken along the $<11\bar{2}0>$ pole corresponding to 12 months aged sample B (a) and 36 month aged sample A (b).

**Figure 3**. FFT images of both HRTEM images of Figure 1. Diffraction patterns are simulated over the FFT image where yellow triangles correspond to InN-w, red and green squares correspond to both InN-zb grains and red and green circles are associated to In2O3 grains. For the latter, the size of the circles is proportional to the intensity.

**Figure 4**. Simulated electron diffraction pattern for InN-zb (a) and In2O3 (b) for the [110] axes. The red arrows are aligned with the growth direction. Position and size of the Bragg mask (blue circle) centered in the spot $(11\bar{\bar{2}})$ referred to In2O3 crystals chosen to construct the moiré fringes mapping (c).

**Figure 5**. Core loss EEL spectra of sample A collected along the z-axis taking as starting point the InN/GaN interface and ending in the InN nanoisland surface

**Figure 6**. Numerical moiré fringes mappings HRTEM image of the nanostructure presented in Figure 1 using the Bragg mask of Figure 4(c). The moiré pattern magnifies lattice and rotation mismatch with respect to a reference vector of the unstrained lattice **g**$_r$.

**Figure 7**. Reconstructed RGB multilayered image for the spatial location of the $In_2O_3$ grains in Figure 2(b). Inverse FFT images from a selected Bragg mask pattern of not sharing $In_2O_3$ spots (red and green smaller circles in Fig. 3(b)) were used for red and green channels. Blue channel is the blank.

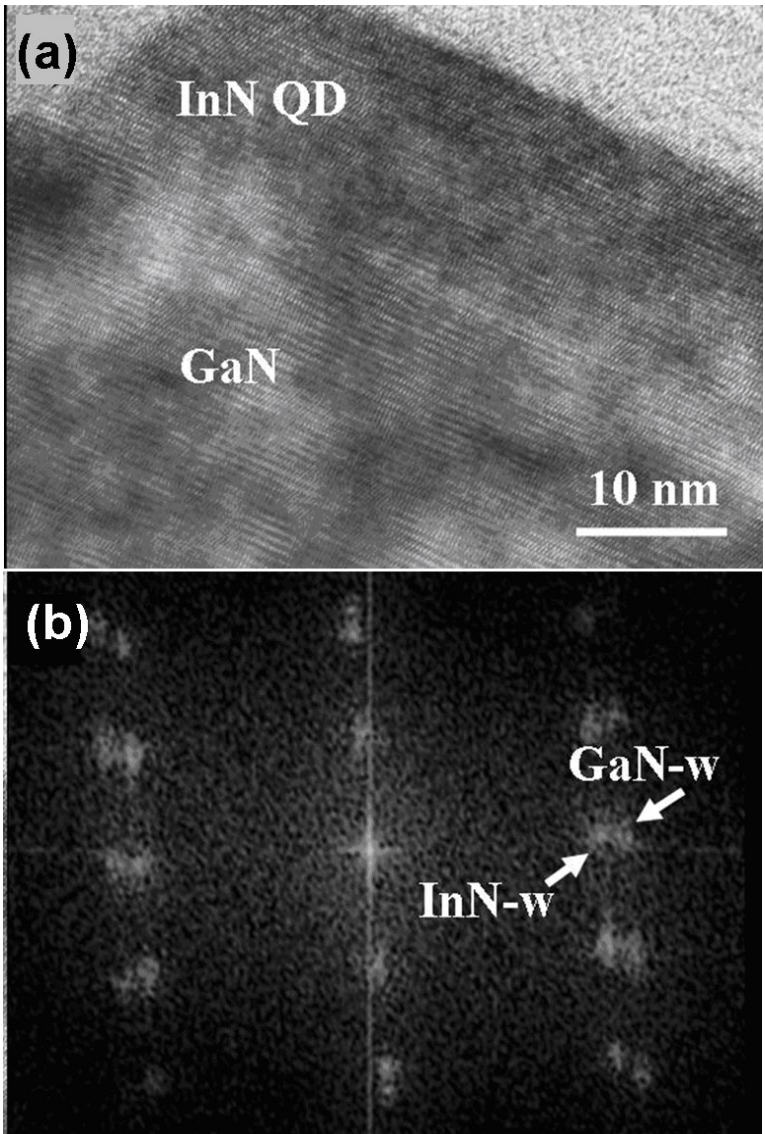

Figure 1

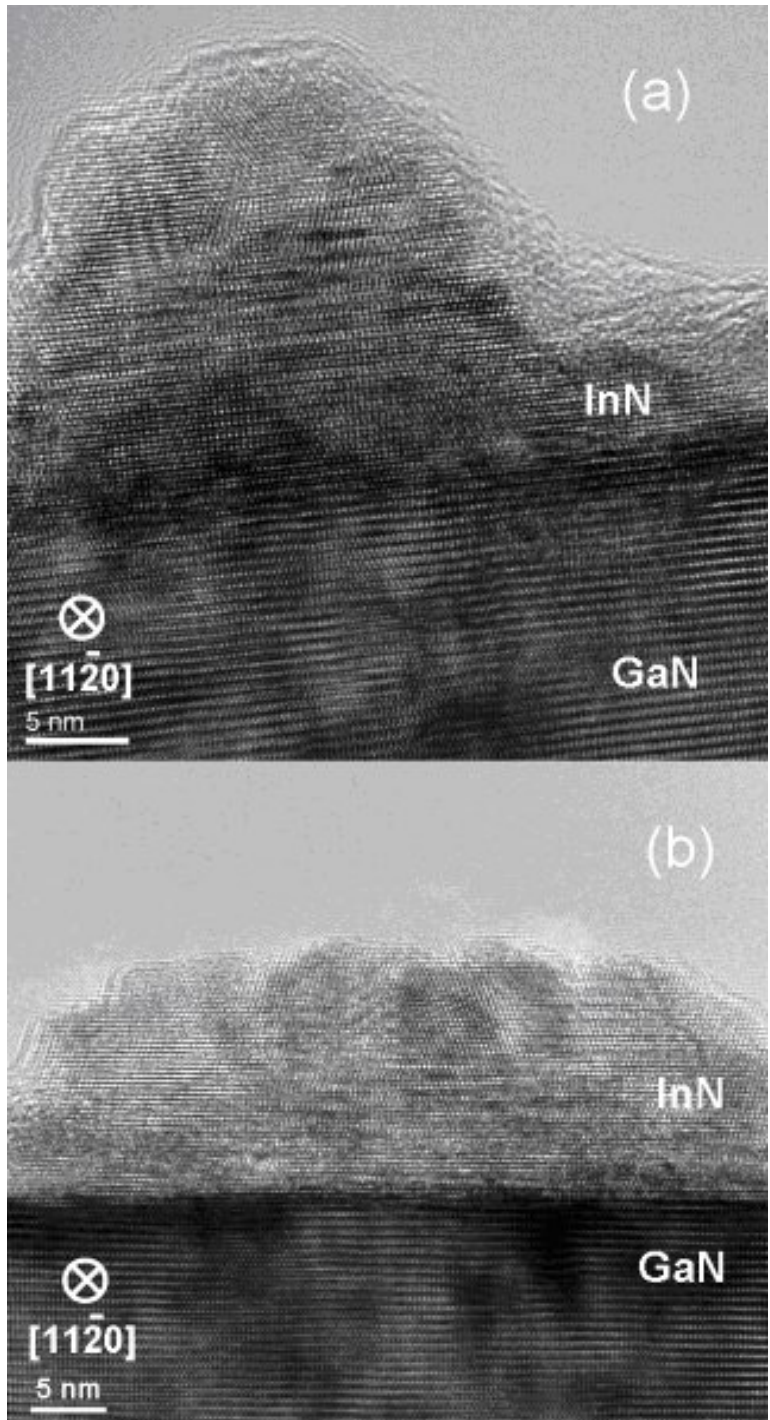

**Figure 2.**

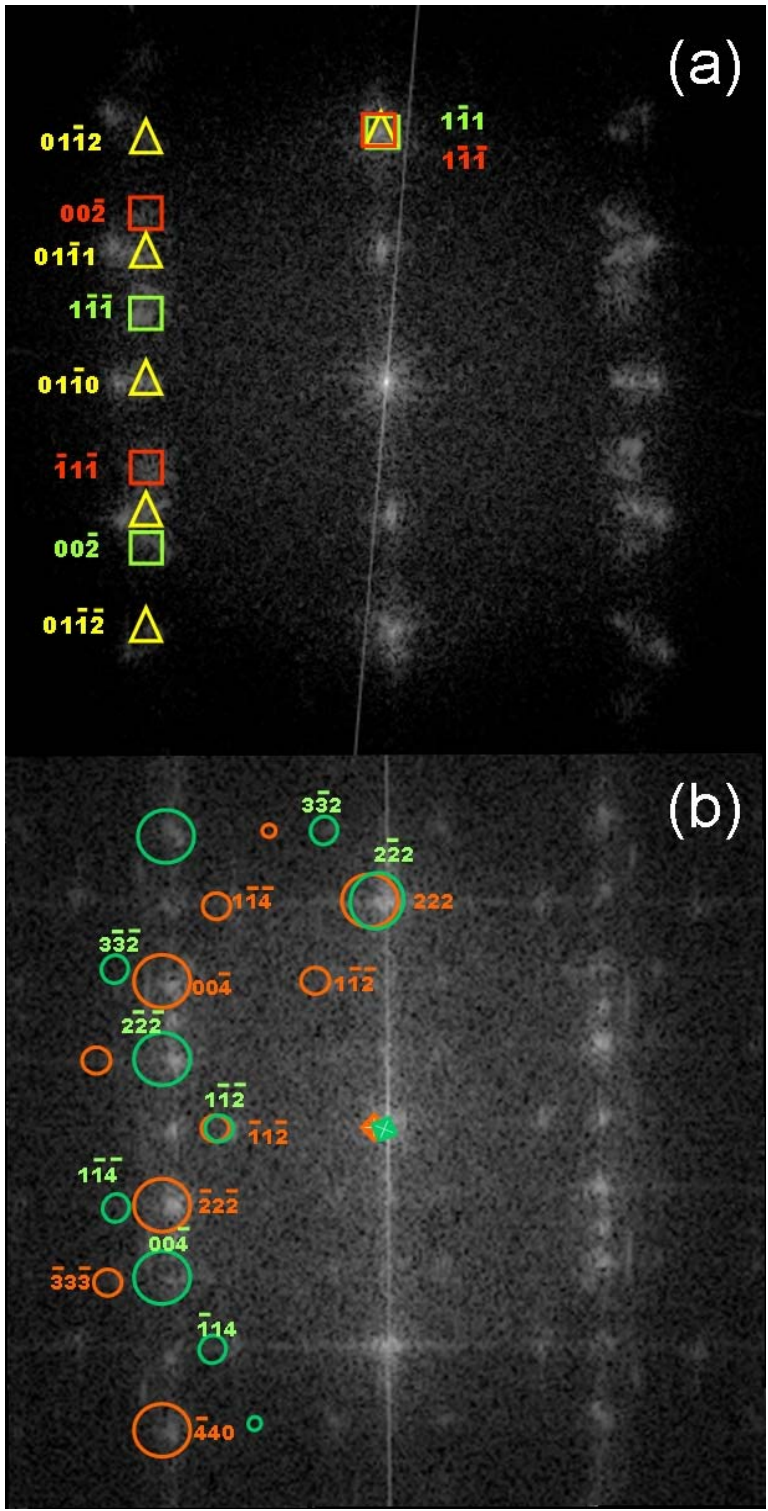

**Figure 3.**

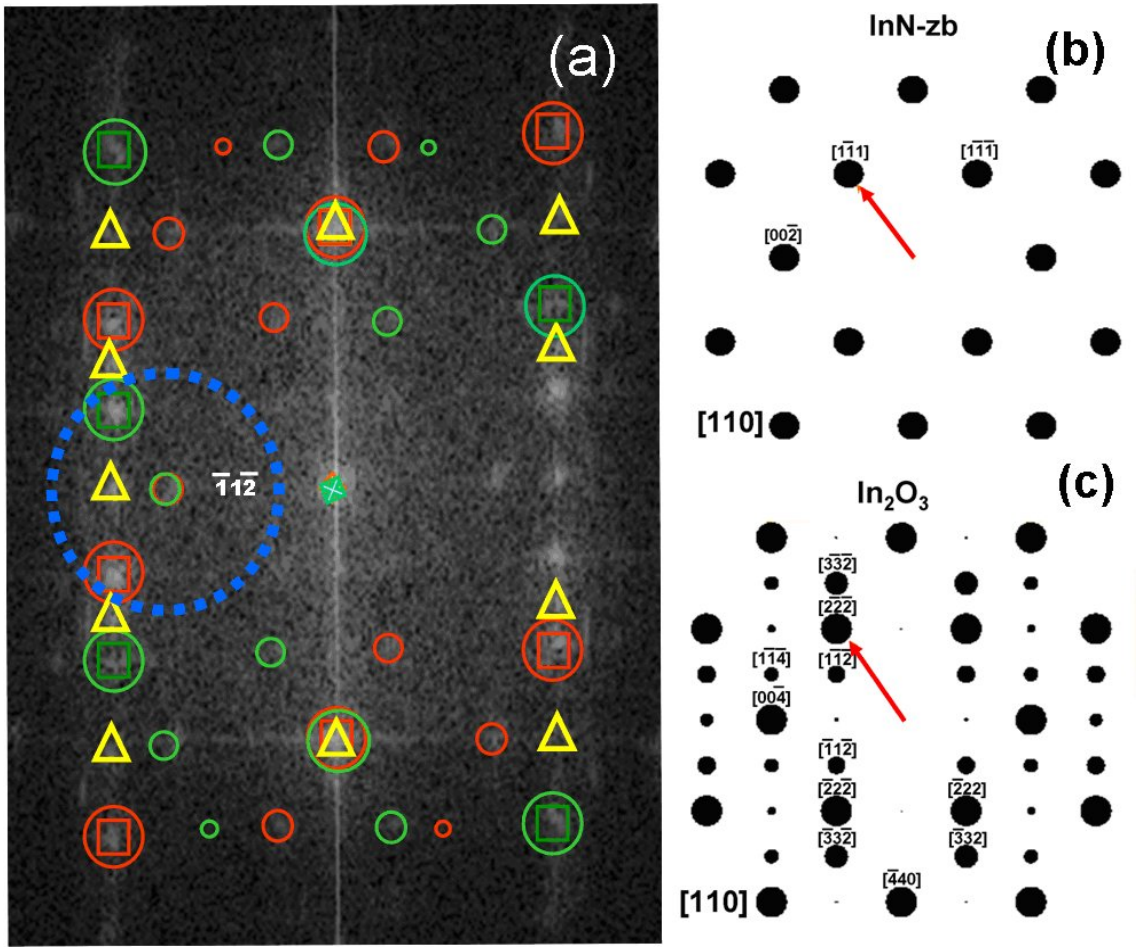

**Figure 4.**

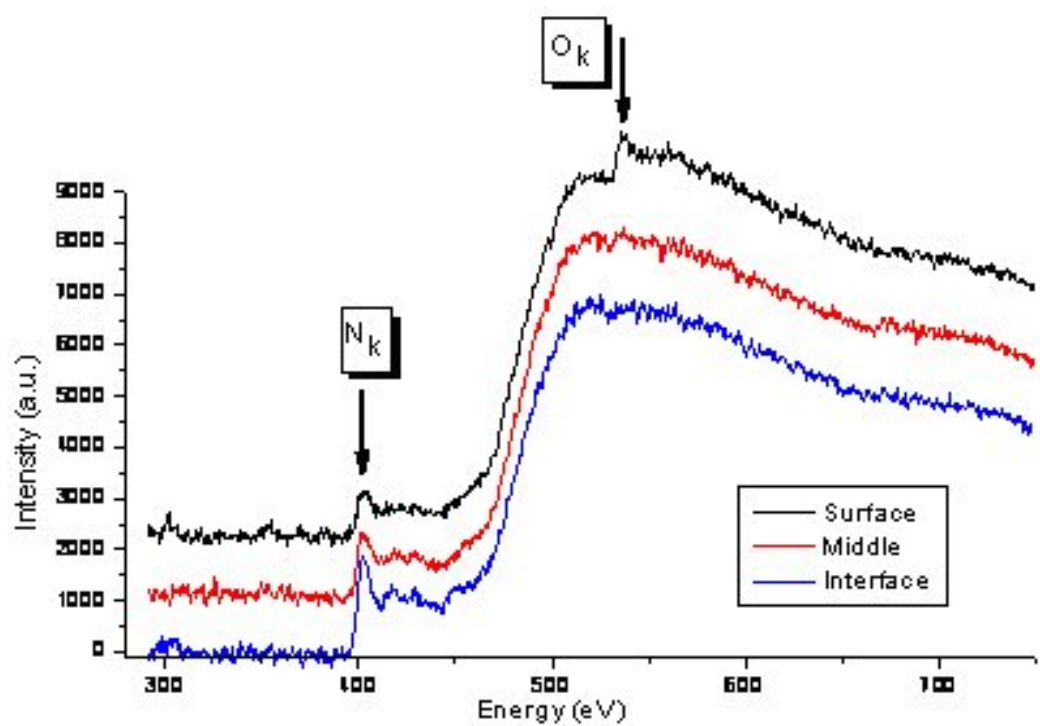

**Figure 5.**

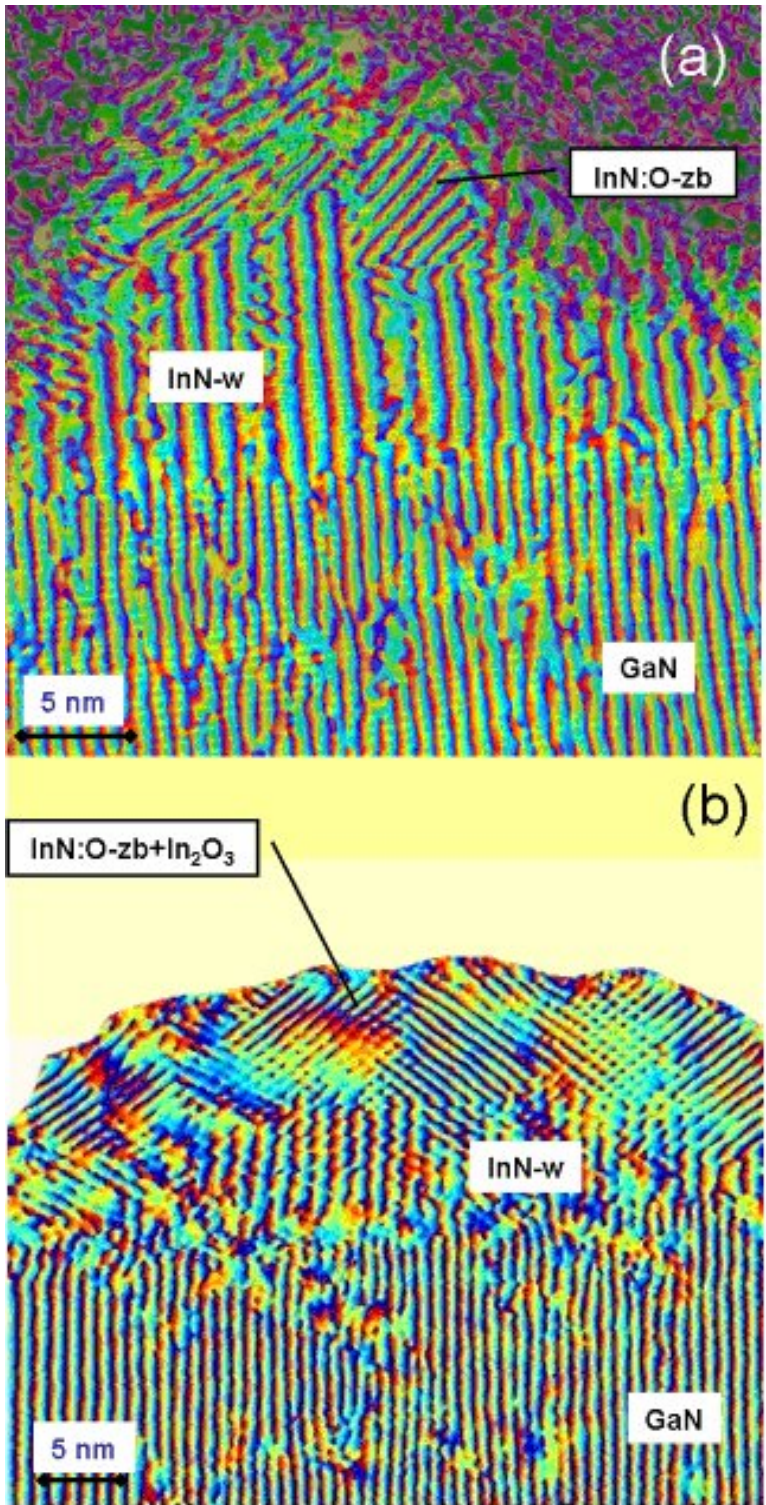

**Figure 6.**

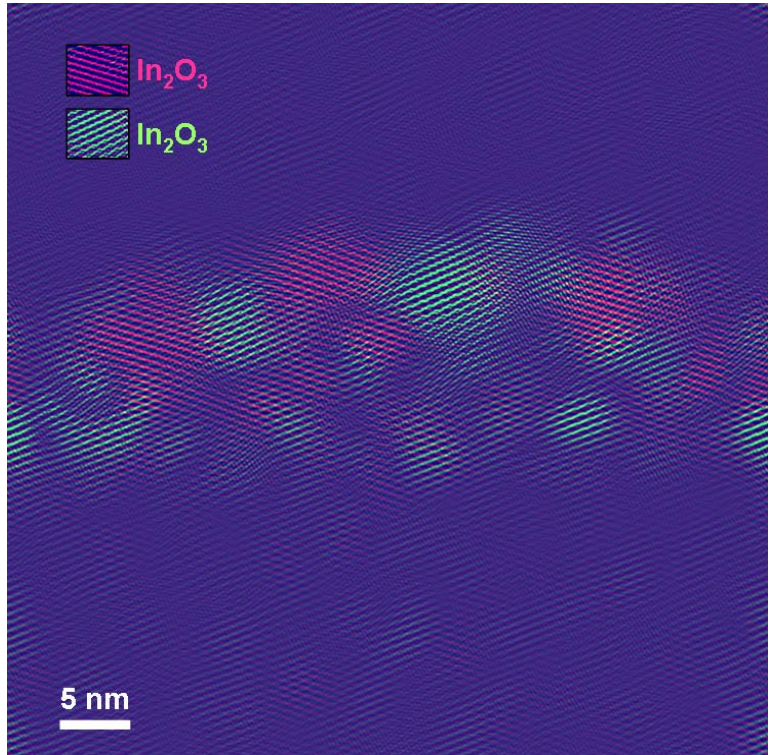

**Figure 7.**